\documentclass[journal=jpccck,manuscript=article]{achemso}

\setkeys{acs}{articletitle=true}

\usepackage[utf8]{inputenc}
\usepackage[english]{babel}
\usepackage{csquotes}
\usepackage{mathpazo}

\usepackage{siunitx}
\usepackage{units}

\usepackage{caption}
\usepackage{float}
\usepackage{setspace}
\usepackage{array}

\usepackage{booktabs}
\usepackage{listings}
\usepackage{lmodern}
\usepackage{microtype}

\usepackage{xcolor}

\usepackage{enumitem}

\usepackage{amsmath,amssymb,amsthm,amstext,amsfonts}

\usepackage{xspace}

\title{NMR study of AgInTe$_{2}$ at normal and high pressure}

\author{Robin Guehne}
\affiliation{%
Felix Bloch Institute for Solid State Physics,
Leipzig University, Linn\'estraße 5, 04103 Leipzig, Germany
}
\email{r.guehne@physik.uni-leipzig.de}
\author{Carsten Kattinger}
\author{Marko Bertmer}

\affiliation{%
Felix Bloch Institute for Solid State Physics,
Leipzig University, Linn\'estraße 5, 04103 Leipzig, Germany
}

\author{Simon~Welzmiller}
\author{Oliver~Oeckler}
\affiliation{%
Institute for Mineralogy, Crystallography and Materials Science, Faculty of Chemistry and Mineralogy, Leipzig University, Scharnhorststr. 20, 04275 Leipzig, Germany
}
\affiliation{%
Institute for Mineralogy, Crystallography and Materials Science, Faculty of Chemistry and Mineralogy, Leipzig University, Scharnhorststr. 20, 04275 Leipzig, Germany
}

\author{Jürgen Haase}%
\affiliation{%
Felix Bloch Institute for Solid State Physics,
Leipzig University, Linn\'estraße 5, 04103 Leipzig, Germany
}

\begin{document}

\begin{abstract}
The ternary semiconductor AgInTe$_2$ is a thermoelectric material with chalcopyrite-type structure that transforms reversibly into a rocksalt-type structure under high pressure. Nuclear magnetic resonance (NMR) is considered to provide unique insight into material properties on interatomic length scales, especially in the context of structural phase transitions. Here, $^{115}$In and $^{125}$Te NMR is used to study AgInTe$_2$ for ambient conditions and pressures up to \SI{5}{GPa}. Magnetic field dependent and magic angle spinning (MAS) experiments of $^{125}$Te prove strongly enhanced internuclear couplings, as well as a distribution of isotropic chemical shifts suggesting a certain degree of cation disorder. The indirect nuclear coupling is smaller for $^{115}$In, as well as the chemical shift distribution in agreement with the crystal structure. The $^{115}$In NMR is further governed by a small quadrupolar interaction  ($\nu_\mathrm{Q} \approx \SI{90}{kHz}$) and shows an orders of magnitude faster nuclear relaxation in comparison to that of $^{125}$Te. At a pressure of about \SI{3}{GPa}, the $^{115}$In quadrupole interaction increases sharply to about \SI{2400}{kHz}, indicating a phase transition to a structure with a well defined, though non-cubic local symmetry, while the $^{115}$In shift suggests no significant changes of the electronic structure. The NMR signal is lost above about $\SI{5}{GPa}$ (at least up to about \SI{10}{GPa}). However, upon releasing the pressure a signal is recovered that points to the reported metastable ambient pressure phase with a high degree of disorder.
\end{abstract}

%%%%%%%%%%%%%%%%%%%%%%%%%%%%%%%%%%%%%%%%%%%%%%%%%%%%

\section{Introduction}

The ternary chalcogenide AgInTe$_2$ is of particular interest as a stoichiometric thermoelectric material for its ability to form solid solutions, e.g. AgIn$_x$Sb$_{1-x}$Te$_2$,\cite{Schroder2013} with a wide range of energy band gaps and lattice parameters.\cite{Kasiviswanathan1986,Welzmiller2014,Welzmiller2015}
AgInTe$_2$ is also known to undergo a structural phase transition from its 4-fold coordinated chalcopyrite-type crystal structure ($I\bar{4}2d$, cf. Fig.~\ref{fig:3}\textbf{a} inset) at ambient conditions to a 6-fold coordinated NaCl-type structure (or possibly a slightly distorted varient thereof) with cation disorder between \SI{3}{} and \SI{4}{GPa}.\cite{Range1969a,Range1969b,Bovornratanaraks2010,Kotmool2015} %according to the pressure-coordination rule\cite{Neuhaus1968}. 
Upon decompression, it has been reported to form a metastable sphalerite-type structure with cation disorder.\cite{Range1969b}

Nuclear magnetic resonance (NMR) studies of this type of material are not known, although, they should provide insight into properties of the various phases from a local probe's perspective. NMR studies of related compounds such as InP date back several years $-$ focusing on chemical properties like bonding character\cite{Lutgemeier1964,Becker1979,Vanderah1988} and nuclear exchange interactions \cite{Shulman1958,Engelsberg1972,Han1988,Tomaselli1998,Adolphi1992,Iijima2004,Iijima2006}$-$ but do not include AgInTe$_2$.

We became interested in this material because of the intriguing behavior under pressure, given our recent endeavor into high pressure NMR studies of solids with anvil cells that currently allow studies under up to about 10 GPa (100 kbar) of pressure.\cite{Haase2009,Meissner2013,Kattinger2021} In order to obtain a high sensitivity with the associated small sample that fits the pressure cell, we use radio-frequency (RF) mircrocoils inside the pressurized region (for a description of the setup see also Fig.~S2~of the Supplementary Information (SI)). In fact, a first application with AgInTe$_2$ appeared to show an insulator-to-metal transition\cite{Meier2015} that we wanted to revisit.

Here we report on a more comprehensive $^{115}$In and $^{125}$Te NMR study of AgInTe$_2$ at ambient and high pressure. With field dependent, MAS, and high pressure NMR we characterize both, chemical properties of the ambient condition phase as well as the structural phase transitions upon pressurization and subsequent pressure relief.

\section{Methods}

AgInTe$_2$ was prepared by melting stoichiometric amounts of elements Ag (\SI{99.999}{\%}, Premion), In (\SI{99.999}{\%}), and Te (puriss., both from VEB Spurenmetalle Freiberg) in a sealed silica ampule under dry Ar atmosphere. After fusing the starting materials at \SI{950}{\celsius}, the ampule was quickly cooled in air, followed by annealing at \SI{400}{\celsius}. The ingots obtained showed gray metallic luster. For the NMR experiments a fine powder was used.

A powder X-ray diffraction pattern (cf. Fig.~S1~in SI) of a representative part of the sample fixed on a specimen holder with Mylar foil was recorded using a Huber G670 Guinier diffractometer with fixed imaging-plate detector and automatic read-out system (Cu-K$_{\alpha1}$ radiation, Ge(111) monochromator, $\lambda=\SI{1.54056}{\AA}$). It showed no impurity, and homogeneity was also confirmed by a Rietveld fit based on a chalcopyrite-type structure model.

NMR experiments were carried out at static magnetic fields of \SI{2.35}{T}, \SI{7}{T}, \SI{9.4}{T}, \SI{11.7}{T}, and \SI{17.6}{T}, as well as with a variable field magnet of up to \SI{16}{T}, using commercial Bruker or Tecmag NMR consoles. Home-built, cryogenic (5 to $\sim$\SI{400}{K}), as well as commercial MAS probes (Bruker) with \SI{4}{mm} rotors were employed. The sample container and the radio frequency (RF) coil were individually optimized for the experiment to allow best excitation conditions and filling factors, except for MAS applications.

We investigated mainly the $^{115}$In nucleus, and obtained some results for $^{125}$Te. 
Single pulse excitation was used to measure the free induction decay (FID) for sufficiently narrow resonance lines or MAS experiments. Spin echo pulse sequences ($\pi/2-\tau-\pi$) were employed to measure broad spectra selectively, as well as transverse relaxation ($T_2$) with different pulse lengths and at different RF power levels. Solid echoes ($\pi/2-\tau-\pi/2$) were employed to excite broad spectra non-selectively. Saturation ($\pi/2-\tau_1-$FID/spin echo) and inversion ($\pi-\tau_1-$FID/spin echo) recovery pulse sequences were used to measure the spin-lattice relaxation ($1/T_1$).

For $^{115}$In, the shift is referenced with respect to In(NO$_3$)$_3$, while for $^{125}$Te, the shift is referenced to Te(CH$_3$)$_2$ (determined by the secondary reference method and metallic copper\cite{Harris2008}).

\paragraph{High Pressure Experiments:} Home-built titanium body anvil cells \cite{Haase2009,Meissner2010} were used for high pressure experiments up to \SI{5}{GPa} (\SI{50}{kbar}) of pressure (the signal was lost at higher pressure). RF microcoils with diameters between \SI{200}{} $\mu$m and \SI{450}{} $\mu$m using insulated silver wire (Goodfellow Cambridge Ltd) with a diameter of 25 $\mu$m (5 $\mu$m insulation) were placed inside the sample chamber (cf. SI). Small amounts of the powder sample were placed within the coil region together with ruby chips for pressure monitoring (optical R1 and R2 line luminescence after laser excitation \cite{Forman1972}). Typical quality factors (Q) of such cells range between 10 to 30 which corresponds to bandwidths of about 3.5 to \SI{11}{MHz} for a $^{115}$In resonance frequency close to \SI{110}{MHz} at \SI{11.7}{T}.

\section{Results}
%%%%%%%%%%%%%%%%%%%%%%%%%%%%%%%%%%%%%%%%%%%%%%%%%
\subsection{Ambient pressure}
\begin{figure}[t]
\centering
\includegraphics[width=.5\textwidth]{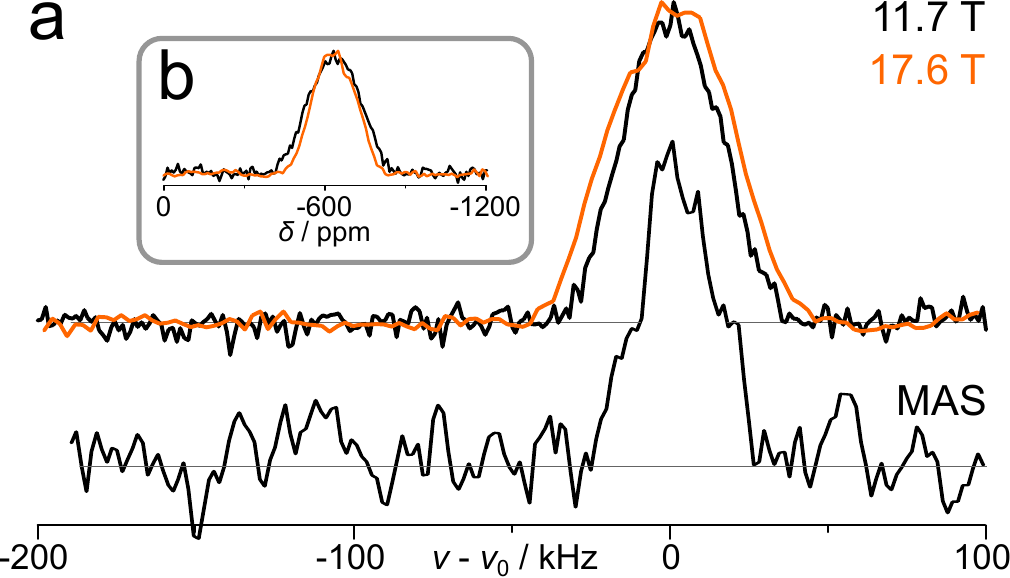}
\caption{\label{fig:1} $^{125}$Te NMR spectra at two different field strengths, \SI{11.7}{} (black) and \SI{17.6}{T} (orange), on the  frequency (\textbf{a}) and ppm (\textbf{b}) scale. The MAS (\SI{12}{kHz}) spectrum (bottom) was recorded at \SI{11.7}{T} and shows hardly any narrowing (no spinning sidebands are seen) when compared to the static spin echo FT spectrum above; at \SI{17.6}{T} the signal broadens only slightly, as the corresponding spectra in (\textbf{b}) show (the spectra in units of ppm are given with respect to Te(CH$_3$)$_2$). For the static measurements we averaged 24 spin echoes with $\pi/2$ pulse lengths of \SI{1.5}{}~$\mu$s and a delay of $\tau=\SI{45}{}$~$\mu$s. For MAS experiments, 4 single pulse FIDs were averaged with \SI{10}{}~$\mu s$ pulse length.}
\end{figure}

%%%%%%%%%%%%%%%%%%%%%%%%%%%%%%%%%%%%%%%%%%%%%%%%%%%%%%
\noindent
We show $^{125}$Te NMR spectra (spin 1/2, $\gamma=\SI{13.545}{MHz/T}$, abundance of about 7\%) in Fig.~\ref{fig:1}. The observed chemical shift of $\SI{-630}{ppm}$ points to Te$^{2-}$,\cite{Morales1997} but we note a significant linewidth of about \SI{250}{ppm} (\SI{39}{kHz} at \SI{11.7}{T}), which does not narrow under sample spinning. Furthermore, the resonance must have a significant field-independent contribution as the linewidth slightly decreases on the ppm scale by going to higher field, cf.~Fig.~\ref{fig:1}{\bf b}. The spin echo decay is rather fast and Gaussian at the beginning, with an approximate $T_{\rm{2G}} \approx \SI{180}{}\ \mu$s (Fig.~S3~in SI), while the spin-lattice relaxation time $T_1$ is found to be about $\SI{360}{s}$. 

%%%%%%%%%%%%%%%%%%%%%%%%%%%%%%%%%%%%%%%%%%%%%%%%%%%%%%%%%%
%%%%%%%%%%%%%%%%%%%%%%%%%%%%%%%%%%%%%%%%%%%%%%%%%%%%%%%%%%

\begin{figure}[t]
\centering
\includegraphics[width=.5\textwidth]{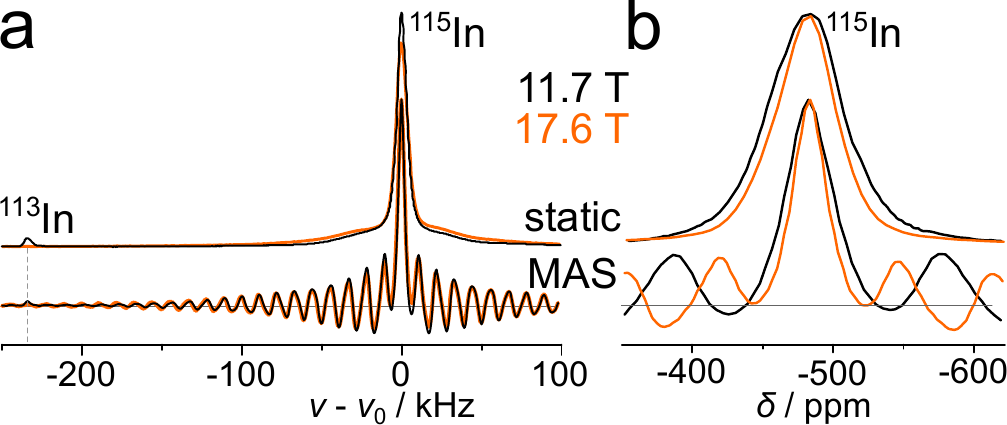}
\caption{\label{fig:2} $^{115}$In NMR spectra at two different magnetic fields  with and without MAS (\SI{10}{kHz}) on ({\bf a}) a large frequency range, and ({\bf b}), the narrow line on the ppm scale. While the central line narrows moderately under rapid spinning, the broad line shows a large number of spinning sidebands. One notes significant field independent contributions to the linewidths. The resonance of the $^{113}$In nucleus can be seen in ({\bf a}) at \SI{-235}{kHz}. The shift is measured with respect to In(NO$_3$)$_3$.}
\end{figure}

\begin{figure}[t]
\centering
\includegraphics[width=.5\textwidth]{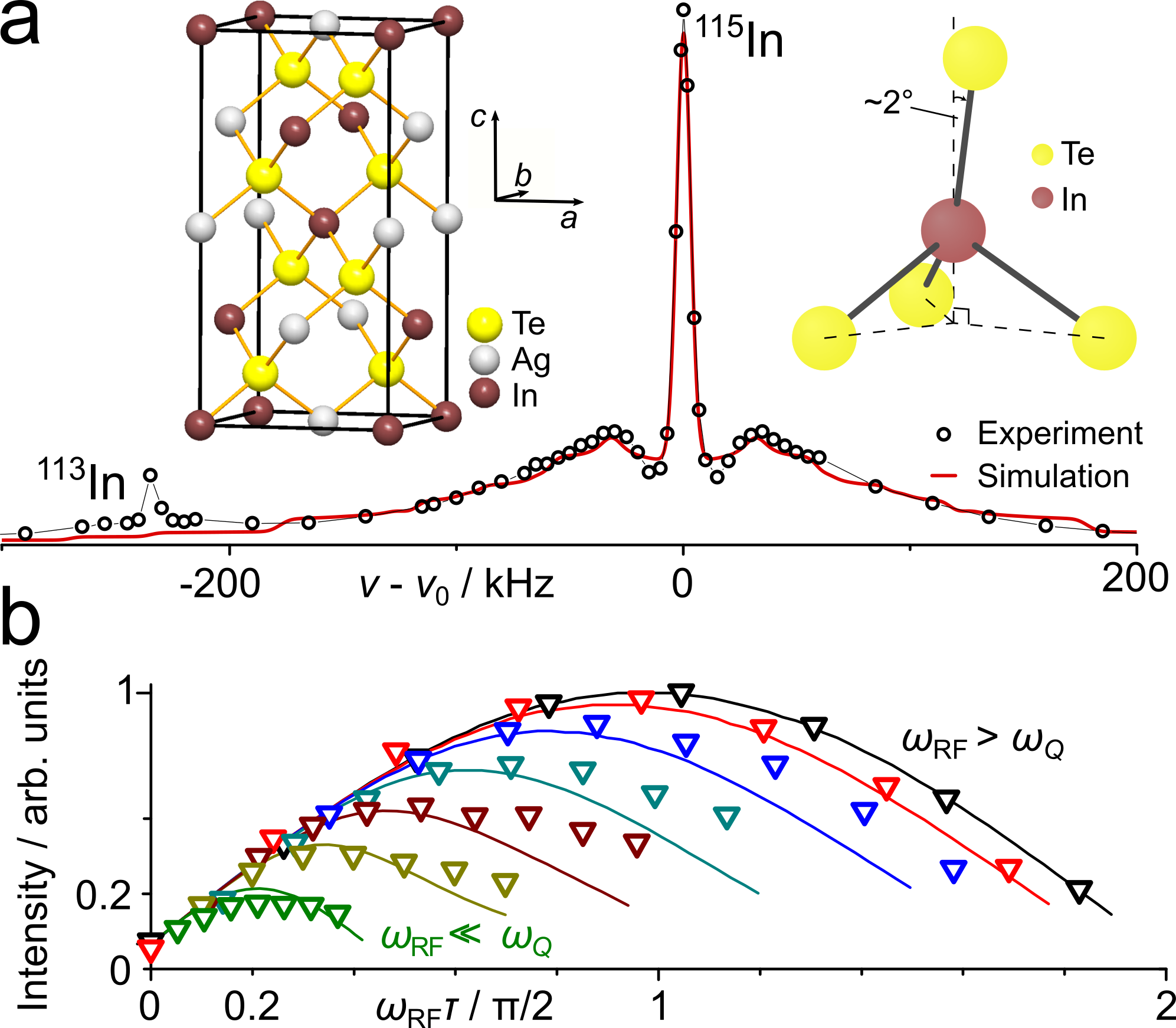}
\caption{\label{fig:3}(\textbf{a}) combined selective spin echoes ($\pi/2$ pulse width \SI{50}{} $\mu$s) of $^{115}$In and $^{113}$In nuclei obtained at \SI{11.7}{T} shown as circles. The red solid line represents a first-order quadrupole powder simulation (see text for details). The inset on the left shows the crystal structure of AgInTe$_2$, and on the right, the arrangement of the 4 Te atoms surrounding In as obtained from Rietveld refinement (cf. SI). (\textbf{b}) nutation experiments of the $^{115}$In central transition (triangles) at different power levels from non-selective excitation with $\omega_{\mathrm{RF}}/\omega_{\mathrm{Q}}\approx4.3$ (black) to selective excitation with $\omega_{\mathrm{RF}}/\omega_{\mathrm{Q}}\approx0.077$ (green). Solid lines represent simulations for various excitation power levels $\omega_{\mathrm{RF}}$.}
\end{figure}

The $^{115}$In NMR spectra (spin 9/2, $\gamma = \SI{9.3857}{MHz/T}$, abundance of about 95\%) are shown in Fig.~\ref{fig:2}. 
The narrow, symmetric peak shows a chemical shift of about \SI{-481}{ppm} with a linewidth of only \SI{62}{ppm} ($\sim$\SI{7}{kHz} at \SI{11,7}{T}). 
The slight deviation of the local symmetry from a perfect cubic environment (cf. inset of Fig.~\ref{fig:3}\textbf{a}) results in a small quadrupole interaction leading to a narrow central transition and a broad resonance given by the orientation dependent satellite transitions of the powder.
Such a satellite distribution is known to average very well under MAS as found in the experiment.
 The spin echo decay is about \SI{90}{}~$\mu $s, though, not a single exponential, and the spin-lattice relaxation is about a factor of  \SI{E4}{} faster (\SI{30}{ms}) than that of Te, again pointing to the importance of quadrupole interaction for the $^{115}$In nucleus ($eQ \approx \SI{77.2(5)E-30}{} \SI{}{|e|m^{-2}}$).
%%%%%%%%%%%%%%%%%%%%%%%%%%%%%%%%%%%%%%%%%%%%%%%%%%%%%%%%%%
%%%%%%%%%%%%%%%%%%%%%%%%%%%%%%%%%%%%%%%%%%%%%%%%%%%%%%%%%%
%%%%%%%%%%%%%%%%%%%%%%%%%%%%%%%%%%%%%%%%%%%%%%%%%%%%%%%%%%
%%%%%%%%%%%%%%%%%%%%%%%%%%%%%%%%%%%%%%%%%%%%%%%%%%%%%%%%%%

In order to better resolve the $^{115}$In NMR signal, we recorded the total spectrum with frequency stepped, selective spin echoes, as well as  broad band quadrupole echoes.\cite{solid}
For the spin echoes we used a probe with a high Q, and long pulses (low power) were employed to ensure selective excitation. A typical spectrum is shown in Fig.~\ref{fig:3}{\bf a}. While the broad peak is much better visible, it does not show singularities typical for powder spectra. Nevertheless, the shape of the broad line with a quite separate and narrow central transition points to a quadrupole tensor with some asymmetry.
We can fit the total spectrum with a quadrupole first-order spectrum with $\nu_Q = \SI{90}{kHz}$, $\eta = 0.3$, and an additional broadening of \SI{7}{kHz}.
Nutation experiments measured of the central transition with a low-Q probe reveal that the broad line is indeed caused by the estimated quadrupole interaction.

\begin{figure}[t]
\centering
\includegraphics[width=.5\textwidth]{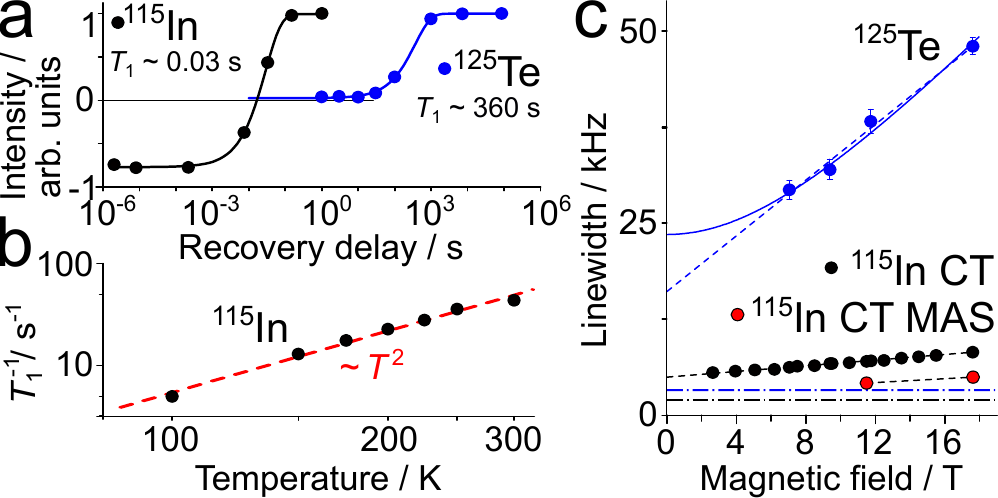}
\caption{\label{fig:4}(\textbf{a}) spin-lattice relaxation measurements of $^{115}$In (black, non-selective excitation, inversion recovery) and $^{125}$Te nuclei (blue, saturation recovery). In panel (\textbf{b}) we show the $^{115}$In relaxation rate as a function of temperature in a log-log plot together with a $T^2$ dependence (red dashed line). 
(\textbf{c}) the static and \SI{10}{kHz} MAS linewidths (full width at half maximum) of the $^{115}$In CT (black and red, respectively), as well as for $^{125}$Te (blue) as functions of the external magnetic field. Dashed lines represent best fits to linear dependences and the solid line a $(\Lambda^2=\Lambda_0^2+(bB_0)^2)$ relationship between linewidth and field. The dash-dotted lines at the bottom denote the calculated line broadening from direct dipole-dipole interactions for $^{115}$In (\SI{2}{kHz}) and $^{125}$Te (\SI{3.3}{kHz}) using the method of moments.\cite{vVleck1948}}
\end{figure}

Finally, we show in Fig.~\ref{fig:4} the results of $^{115}$In and $^{125}$Te relaxation measurements, including a temperature dependence of the $^{115}$In spin-lattice relaxation rate that verifies the  interaction with phonons to be the main cause. Also shown are the field dependences of the linewidths of $^{115}$In and $^{125}$Te to determine possible field-independent contributions. The results are discussed below.

%%%%%%%%%%%%%%%%%%%%%%%%%%%%%%%%%%%%%%%%%%%%%%%%%
\subsection{Elevated pressure}
\begin{figure}[t]
\centering
\includegraphics[width=.5\textwidth]{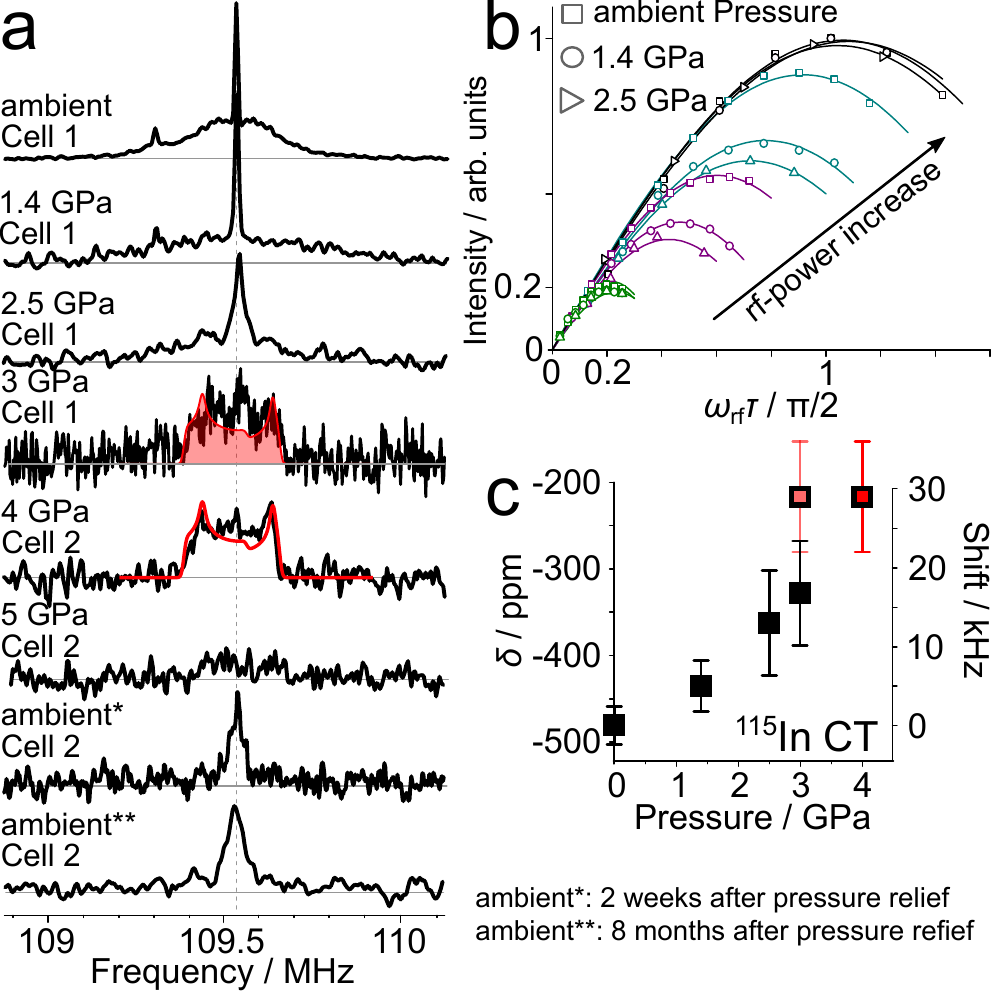}
\caption{\label{fig:5}({\bf a}) Spectra at various pressures including ambient pressure after pressure relief (($^{\ast}$)~and ($^{\ast\ast}$)) at the bottom of the panel. Quadrupole echoes (\SI{0.5}{}$\mu$s $\pi/2$ pulses) were used for recording, except for the spectra at \SI{4}{GPa} and \SI{5}{GPa}, for which spin echo sequences with \SI{0.5}{} - \SI{1.0}{}$\mu$s were used. Spectra were normalized since the signal-to-noise varied. The red shaded area in the \SI{3}{GPa} spectrum represents a second-order quadrupole simulation as obtained from the \SI{4}{GPa} spectrum (red solid line, see text for details). ({\bf b}) nutation experiments (colored symbols) for ambient, \SI{1.4}{} and \SI{2.5}{GPa} pressure confirm a moderate increase of the quadrupole interaction with increasing pressure. Data points  with the same color denote the same excitation power levels. Solid lines are guides to the eye. ({\bf c}) magnetic shift of the central transition as a function of pressure, in ppm with respect to In(NO$_3$)$_3$, and in kHz with respect to the ambient pressure resonance frequency. The 3 and \SI{4}{GPa} values (red data point) are corrected for the second-order quadrupole shift.}
\end{figure}

%%%%%%%%%%%%%%%%%%%%%%%%%%%%%%%%%%%%%%%%%%%%%%%%%%%%%%
As described under \emph{Methods}, NMR anvil cells with RF microcoils were used for these experiments. The very small coil volume and the inherently low Q of the coils warrant high time resolution, but also limit the signal-to-noise. Therefore, we only report on the $^{115}$In resonance, mostly recorded after non-selective quadrupole echo excitation.\cite{solid} The results are summarized in Fig.~\ref{fig:5}.

%%%%%%%%%%%%%%%%%%%%%%%%%%%%%%%%%%%%%%%%%%%%%%%%%%%
%%%%%%%%%%%%%%%%%%%%%%%%%%%%%%%%%%%%%%%%%%%%%%%%%%%

%With respect to Fig.~\ref{fig:5}{\bf a} we note the following. 
At ambient pressure (Fig.~\ref{fig:5}{\bf a}) we obtain within error the same spectrum as in Fig.~\ref{fig:3}{\bf a}. The change of shape of the broad foot as the pressure increases to \SI{1.4}{} and \SI{2.5}{GPa} is due to an increase in the quadrupole frequency, which is proven with the nutation experiments shown in panel ({\bf b}).
At \SI{3}{GPa}, the narrow signal is nearly lost, and replaced by a second-order central line quadrupole powder pattern (red shaded area), while some portion of the nuclear spins seem still to be subject to a smaller quadrupole interaction as they form the rather narrow peak in the middle of the spectrum. 
At \SI{4}{GPa}, the second-order quadrupole spectrum is much better resolved. A fit with a $\nu_Q=\SI{2420}{kHz}$ and $\eta=0.18$ is shown (the same as used for the red shaded area in the \SI{3}{GPa} spectrum). Unfortunately, above \SI{5}{GPa} we lost the NMR signal. Then we released the pressure and measured again at ambient condition to find the spectra (solid echo FT) shown at the bottom in panel ({\bf a}) (acquired about two weeks ($^{\ast}$) and eight months ($^{\ast\ast}$) after decompression). We find a resonance line at the ambient pressure shift but with a much larger linewidth of about \SI{50}{kHz}.

The chemical shift of the central transition as a function of pressure shows a significant increase (from \SI{-481}{} to about \SI{-220}{ppm}), but not such a sudden change as for the quadrupole interaction.

%%%%%%%%%%%%%%%%%%%%%%%%%%%%%%%%%%%
%%%%%%%%%%%%%%%%%%%%%%%%%%%%%%%%%%%
\section{Discussion}
%%%%%%%%%%%%%%%%%%%%%%%%%%%%%%%%%%%
%%%%%%%%%%%%%%%%%%%%%%%%%%%%%%%%%%%
\subsection{Ambient pressure}
The rather long $^{125}$Te (spin-1/2) spin-lattice relaxation time, $T_1 \approx \SI{6}{min}$, is perhaps expected for this semiconducting material with a sizable bandgap and shows that magnetic excitations are largely absent. The negative shift is expected for a Te$^{2-}$ ion, but the large $^{125}$Te NMR linewidth ($\approx \SI{250}{ppm}$) points to disorder in the Te environment. A certain degree of disorder in the Ag and In distribution, which cannot be elucidated in a straightforward way by X-ray crystallography due to the lacking scattering contrast, is a likely explanation. For example, if Te has 3 In and 1 Ag as nearest neighbors (instead of 2 In and 2 Ag), the tetrahedral environment will be more strongly distorted, which likely affects more the isotropic shift. 
Four In or 4 Ag neighbors would result in different isotropic shifts. Therefore, such disorder, if abundant, should contribute to the linewidth. And since we detect roughly all $^{125}$Te nuclei, this would explain the observed shift variation and the rather Gaussian lineshape that does not narrow under MAS. For a more quantitative study, compounds with Ag$_3$In and AgIn$_3$ environments for Te would be desirable, as well as first principle calculations. The spin echo decay of $^{125}$Te is Gaussian at the beginning and is likely to be given by the $^{125}$Te - $^{125}$Te homonuclear interaction. 
The strong field independent linewidth (up to about \SI{25}{kHz}, cf. Fig.~\ref{fig:4}\textbf{c}) points to a large internuclear coupling ($J$-coupling) predominately between $^{125}$Te nulcei and their $^{115}$In neighbors, which can also contribute to the linewidth under MAS. Such enhanced internuclear interactions involving heavy element nuclei have also been reported for related semiconductors.\cite{Shulman1958,Engelsberg1972,Han1988,Adolphi1992,Tomaselli1998,Iijima2004,Iijima2006} 

The nuclear relaxation of $^{115}$In is about 4 orders of magnitude faster than that of $^{125}$Te. Since of the gyromagnetic ratios ($^{115}\gamma/^{125}\gamma \approx 0.69$) cannot account for such a difference in $T_1$, this points to a strong quadrupolar mechanism. The $^{115}$In $T_1$ temperature dependence shown in Fig.~\ref{fig:3}\textbf{b} (quadratic in temperature above half of the Debye temperature, $\theta_{\mathrm{D}}=\SI{202}{K}$\cite{Aikebaier2012}) documents that Raman processes involving all phonons cause the observed behavior.\cite{Kranendonk1954,Kranendonk1967} It might even be possible to relate the relaxation rate to the actual density of states of the phonons if one could compare with corresponding data on different systems. 
The observed quadrupole splitting due to the slightly distorted tetrahedral environment is not well resolved and the linewidths vary slightly between the frequency stepped spin echo data and the Fourier transform of quadrupole echoes (with long pulse delays to suppress the influence of Solomon echoes\cite{solid}). Also, one notices the absence of any singularities often found with powder patterns. This points to distributions of the tensor components of the electric field gradient, but given the mentioned uncertainties we did not attempt to pursue particular fits. We estimate a quadrupole frequency of about  \SI{90}{kHz} with an asymmetry parameter $\eta \approx 0.3$. 

The $^{115}$In NMR line at about \SI{-480}{ppm} has a linewidth of about \SI{62}{ppm} that narrows under MAS to about \SI{28}{ppm} ($\sim$\SI{3}{kHz}). As Fig.~\ref{fig:4}{\bf c} shows, the field dependent contributions to the $^{115}$In linewidth are much smaller than for $^{125}$Te. This is expected since all $^{115}$In neighbors are Te atoms, so that a somewhat random occupation of the cation sites is less important. Due to the low abundance of $^{125}$Te, the local field at $^{115}$In is mostly given by other In nuclei in the second shell (Ag has low-$\gamma$ nuclei and does not contribute significantly). Therefore, the indirect coupling that contributes to the field independent linewidth (about \SI{5}{kHz}) is expected to be much weaker at $^{115}$In, as well. The spin echo decays can be found in the SI.

\subsection{Pressure dependence}

The dependence of the quadrupole coupling on pressure is interesting. After a moderate change of the quadrupole splitting up to about \SI{2.5}{GPa}, a sudden increase at \SI{3.0}{GPa} follows. Importantly, this phase has a rather well-defined second-order quadrupole pattern, pointing to a more or less well ordered structure. 
Certainly, this finding disagrees with a cubic NaCl structure that was discussed in the literature, because cation disorder (second coordination sphere) is not expected to yield such a strong and well defined deformation of the local charge distribution. Rather, our data point at a crystal structure with a non-cubic In environment. Based on the diffraction of synchrotron radiation of AgInTe$_2$ for pressures up to \SI{6.2}{GPa}, Bovornratanaraks et al. (2010) come to a similar conclusion.\cite{Bovornratanaraks2010}
While they observe slight changes of the reflection patterns for low pressures and a gradual replacement of the ambient structure by the high pressure phase around \SI{3}{} to \SI{4}{GPa} (in good agreement with our quadrupole data), the authors question the consensus of a NaCl-type high pressure structure in favor of a cation-disordered $Cmcm$ phase, although XRD data remain ambiguous. 
Here, NMR provides additional evidence from a local perspective for such a crystal structure with non-cubic $^{115}$In environment. 
Aided by first principles calculations, the quadrupole frequency and the asymmetry parameter as obtained from the second-order quadrupole fit may by used to assess the local In environment of the high pressure phase in greater detail.
Finally, the rather moderate changes of the chemical shift of about \SI{150}{ppm} by going form the chalcopyrite-type to the high pressure structure do not support significant changes in the electronic structure and thus, the occurrence of a metallic phase.

Unfortunately, we lost the signal above \SI{4}{GPa}. With various pressure cells we searched for signal in a broad range of frequencies up to about \SI{10}{GPa} since we expected a metallic phase.\cite{Meier2015} Measurements after pressure relief, however, prove a proper functioning of the anvil cells and the loss of signal must therefore have physical reasons. What caused the disappearance of the signal cannot be said with certainty as these measurements are difficult and time consuming. Possible reasons are further broadening of the NMR signal or an increase in relaxation time.  

After we released the pressure from one of the cells, we obtained a spectrum governed by a single, rather narrow peak at the ambient pressure resonance frequency (cf. Fig.~\ref{fig:5}~(a), bottom spectrum), which has not noticeably changed after about 8 months under ambient conditions (the better signal-to-noise ratio of the second measurement is due to a greater number of averaged scans).  
The broadening of this resonance line ($\sim\SI{50}{kHz}$), however, is much larger in comparison to the initial linewidth for ambient pressure and the spectra may thus represent the metastable phase with zincblende-type structure (AgInTe$_2$-III) and randomly distributed cations.\cite{Range1969a,Range1969b}

The high pressure NMR data prove that the various stages of the phase transitions in a material like AgInTe$_2$ under the application of external pressure can clearly be distinguished, and using an optimized high pressure setup in combination with first principles calculations, structural as well as electronic properties could be studied with local resolution and high accuracy.

\section{Conclusions}

We investigated the $^{115}$In and $^{125}$Te NMR of powdered AgInTe$_2$ at ambient and $^{115}$In NMR at high pressures. 
The $^{125}$Te linewidth, studied for various fields and under MAS, suggests strongly enhanced internuclear interactions, as well as a distribution of isotropic shifts, yielding a large field independent linewidth and weak narrowing under MAS. Both features hold for the $^{115}$In nuclei, though less pronounced, in agreement with a more symmetric local In environment. 
The internuclear coupling very likely causes the rather rapid and complex transverse relaxation phenomenology observed for both nuclei. Furthermore, $^{115}$In nuclei are subject to a small quadrupole interaction of about \SI{90}{kHz} and feature a quadrupolar spin-lattice relaxation orders of magnitude faster compared to that of $^{125}$Te nuclei. With increasing pressure, the quadrupole interaction increases moderately at the beginning, but sharply around \SI{3}{GPa} to about \SI{2400}{kHz}. This indicates a phase transition to a rather well ordered structure with lowered local symmetry, in agreement with literature data. The shift increases almost continuously from about \SI{-500}{} to \SI{-200}{} in the same pressure range, implying no significant electronic changes. Above about \SI{5}{GPa}, the NMR signal is lost and could not be found up to about \SI{10}{GPa}. Upon decompression, however, we recovered a resonance line with the same shift but a much bigger linewidth as the initial ambient spectrum. The line broadening may be due to the reported\cite{Range1969b} metastable sphalerite-type structure with random cation distribution.

\section{Supporting Information}

\section{Powder X-ray diffraction and Rietveld refinement}

\begin{figure}[h!]
\centering
\includegraphics[width=\textwidth]{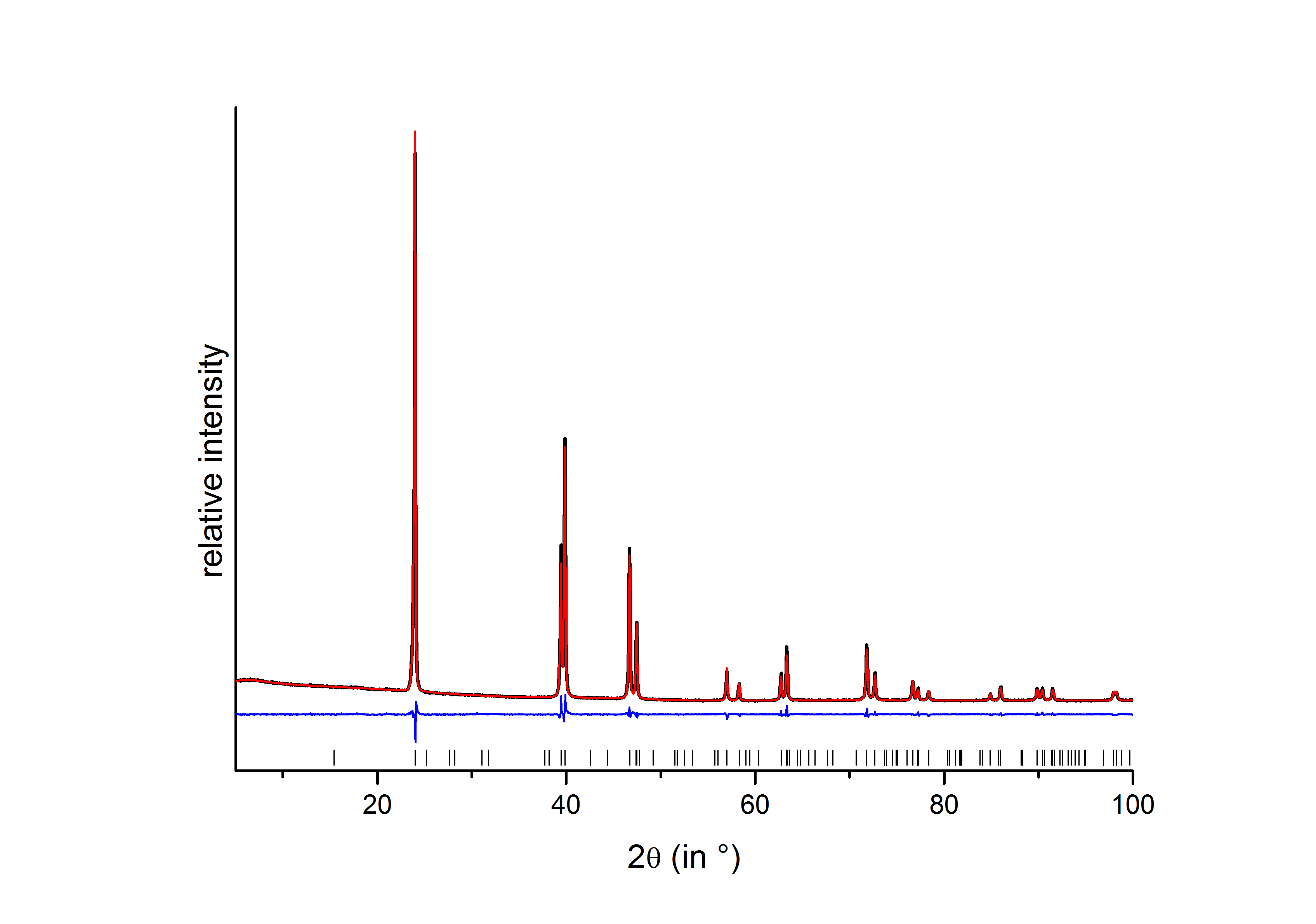}
\caption{Black, powder XRD reflection pattern of AgInTe$_2$ as measured, in Red, results from the Rietveld refinement, and in Blue, the difference plot. Rietveld refinement confirms space group $I\bar{4}2d$ and determines $a=6.45741(6)$ and $c= 12.65146(17)$.}
\end{figure}

\noindent
A Huber G670 Guinier camera equipped with a fixed imaging-plate and an automatic read-out system (CuK$_{\alpha1}$ radiation, Ge(111) monochromator, $\lambda=1.54056$ \AA) was used to record the powder X-ray diffraction patterns of the investigated sample to quantify sample homogeneity and lattice parameters. To fix representative parts of the powdered sample on a specimen holder with Mylar foil, hair-fixing spray was used.

Using Topas Academic software, Rietveld and Pawley fits were carried out. The reflection profiles were described with a direct convolution approach and fundamental parameters. The emission profile was described using a Pseudo-Voigt function.

\begin{table}
\centering
\caption{Rietveld refinement details}
\begin{tabular}{lllll}
\toprule
R$_{\mathrm{p}}$ & R$_{\mathrm{wp}}$ & R$_{\mathrm{wp}}^{\ast}$ & R$_{\mathrm{Bragg}}$ & $\chi^2$\\
0.0177 & 0.0277 & 0.0834 & 0.0169 & 1.232\\
\bottomrule
\multicolumn{5}{l}{$^{\ast}$background subtracted}
\end{tabular}
\end{table}

\section{High-pressure NMR setup}

\begin{figure}[h!]
\centering
{\LARGE{\textbf{\textsf{a}}}}\hspace{0.1cm}\includegraphics[scale=0.3]{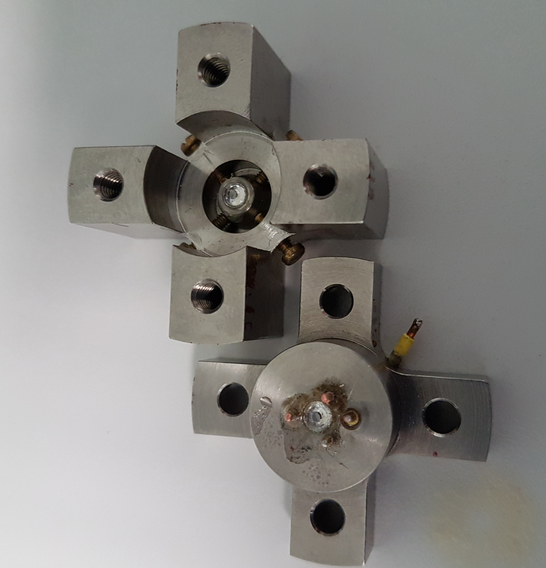}
\hspace{0.5cm}
{\LARGE{\textbf{\textsf{b}}}}\includegraphics[scale=0.7]{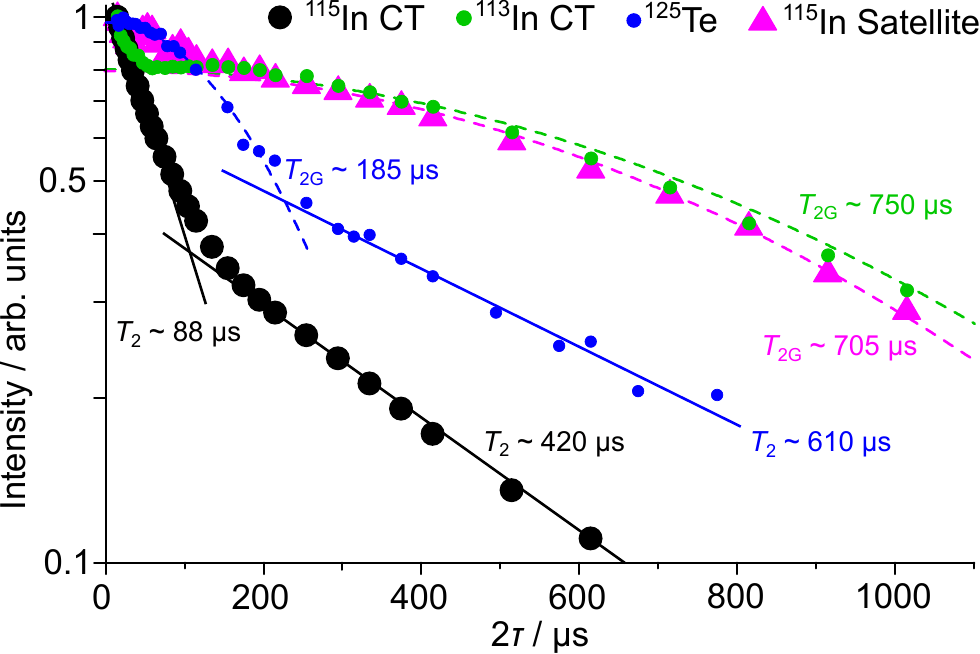}
\caption{High pressure NMR setup. (a) The open titanium cell as used in high pressure experiments, with the two Moissanite visible in the middle of the two cell halves. In the closed cell, the tips (culets) of the two anvils enclose the sample chamber as depicted in panel (b), containing the NMR microcoil with the sample, ruby chips for pressure monitoring, as well as a pressure medium (paraffin oil) to ensure hydrostatic conditions. A non-magnetic CuBe gasket keeps the arrangement in place.}
\end{figure}

\section{Transverse relaxation measurements}

\begin{figure}[h!]
\centering
\includegraphics[scale=1]{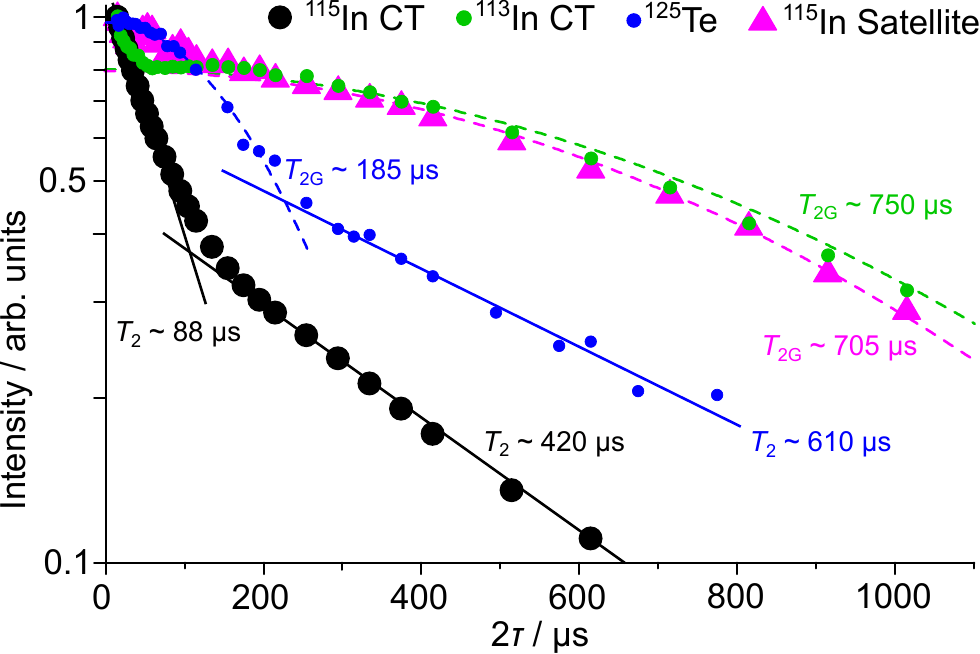}
\caption{\label{FigS4}Spin echo decay measurements obtained for various resonances. All measurements were carried out under selective excitation conditions. '$^{115}$In satellite' represents a narrow slice in the quadrupole powder spectrum about 150 kHz above the central transition where it was ensured that the latter is not excited during the measurement. Solid lines represent single exponential decays ($T_2$), while dashed lines assume Gaussian decays ($T_{2\textrm{G}}$).}
\end{figure}

In Fig.\ref{FigS4}, we present the spin echo decays observed for both In nuclei and Te. The decay of the selectively excited central transition for $^{115}$In is very fast ($T_2\approx\SI{88}{}$ $\mu$s) at the beginning before the slope changes to a much slower decay with $T_2\approx\SI{420}{}$ $\mu$s. Both decays are well approximated with a single exponential. Similar experiments on the satellites give even larger values of $T_{2\mathrm{G}}\approx \SI{705}{}$~$\mu$s, while the shape of the decay is Gaussian-like. This means, the spin flips allowed for the central levels (not affected by the quadrupole interaction) must be rather effective, while being quenched in the satellites \cite{Haase1993b}. This is also borne out by the $^{113}$In NMR measurements. This low abundance nucleus has essentially $^{115}$In neighbors due to the low abundance of $^{125}$Te and the negligible Ag magnetic moments, and we further can neglect any like-spin interactions. It thus measures the local field fluctuations from surrounding $^{115}$In spins, part of which will be in the central levels and show rapid spin exchange, for others it will be quenched. As a consequence we observe a two component decay also for the $^{113}$In CT, for which the rapid decay stems from the effective homonuclear interaction that affects a refocussing of the second pulse. For $^{125}$Te, as expected, we find a similar behavior with a rather fast, Gaussian spin echo decay at the beginning followed by a slower decay for $\tau$ greater than about \SI{125}{} $\mu$s.
 
These spin echo decay data highlight the complexity of the present couplings among the various nuclear magnetic moments. Although we are not able to provide quantitative description of this behavior, the mere rapidness of the observed decays, e.g., about \SI{90}{} $\mu$s and \SI{185}{} $\mu$s for the $^{115}$In CT and $^{125}$Te, respectively, strongly hints at enhanced dipolar interactions. Such strong internuclear interactions would then explain the large field independent linewidths as well as the failure of MAS to lead to a significant narrowing in particular for $^{125}$Te, cf. Fig.1 of the main manuscript. 
For $^{115}$In the residual linewidth of 4 to \SI{5}{kHz} is in agreement with the effective homonuclear decay of the central transition, Fig.4\textbf{b} in the main manuscript. 
Faster spinning should be more effective. However, we are also not certain whether the enhanced coupling is of pseudo dipolar origin and thus shows an angular dependence under rotation. 
It might as well be a result of an indirect nuclear coupling of the Bloembergen-Rowland type %\cite{BR1955} 
with a fairly isotropic line broadening such that it would not be affected by MAS. 
If aided by first principle calculations, more information about the details for the structure could be readily obtained. 
Such enhanced internuclear interactions that involve electronic exchange couplings and yield effective line broadening even under MAS have already been reported for similar systems, namely for InP, InAs, and InSb\cite{Shulman1958,Engelsberg1972,Han1988,Adolphi1992,Tomaselli1998,Iijima2004,Iijima2006}.

%\bibliographystyle{unsrt}
%\bibliography{references}

\section{Acknowledgement}

We thank V. Chlan (Prague) for fruitful discussions and G. Klotzsche (Leipzig) for technical support. We acknowledge
the financial support by the Deutsche Forschungsgemeinschaft, projects 317319632 and 442459148, and by Leipzig University.

\bibliography{references}

\providecommand{\latin}[1]{#1}
\makeatletter
\providecommand{\doi}
  {\begingroup\let\do\@makeother\dospecials
  \catcode`\{=1 \catcode`\}=2 \doi@aux}
\providecommand{\doi@aux}[1]{\endgroup\texttt{#1}}
\makeatother
\providecommand*\mcitethebibliography{\thebibliography}
\csname @ifundefined\endcsname{endmcitethebibliography}
  {\let\endmcitethebibliography\endthebibliography}{}
\begin{mcitethebibliography}{33}
\providecommand*\natexlab[1]{#1}
\providecommand*\mciteSetBstSublistMode[1]{}
\providecommand*\mciteSetBstMaxWidthForm[2]{}
\providecommand*\mciteBstWouldAddEndPuncttrue
  {\def\EndOfBibitem{\unskip.}}
\providecommand*\mciteBstWouldAddEndPunctfalse
  {\let\EndOfBibitem\relax}
\providecommand*\mciteSetBstMidEndSepPunct[3]{}
\providecommand*\mciteSetBstSublistLabelBeginEnd[3]{}
\providecommand*\EndOfBibitem{}
\mciteSetBstSublistMode{f}
\mciteSetBstMaxWidthForm{subitem}{(\alph{mcitesubitemcount})}
\mciteSetBstSublistLabelBeginEnd
  {\mcitemaxwidthsubitemform\space}
  {\relax}
  {\relax}

\bibitem[Schr\"{o}der \latin{et~al.}(2013)Schr\"{o}der, Rosenthal, Souchay,
  Petermayer, Grott, Scheidt, Gold, Scherer, and Oeckler]{Schroder2013}
Schr\"{o}der,~T.; Rosenthal,~T.; Souchay,~D.; Petermayer,~C.; Grott,~S.;
  Scheidt,~E.-W.; Gold,~C.; Scherer,~W.; Oeckler,~O. A high-pressure route to
  thermoelectrics with low thermal conductivity: The solid solution series
  AgIn$_x$Sb$_{1-x}$Te$_2$ ($x=0.1–0.6$). \emph{J. Solid State Chem.}
  \textbf{2013}, \emph{206}, 20--26\relax
\mciteBstWouldAddEndPuncttrue
\mciteSetBstMidEndSepPunct{\mcitedefaultmidpunct}
{\mcitedefaultendpunct}{\mcitedefaultseppunct}\relax
\EndOfBibitem
\bibitem[Kasiviswanathan \latin{et~al.}(1986)Kasiviswanathan, Rao, and
  Gopalam]{Kasiviswanathan1986}
Kasiviswanathan,~S.; Rao,~D.~K.; Gopalam,~B. Preparation and characterization
  of the pseudobinary system Cu$_{1-x}$Ag$_x$InTe$_2$. \emph{J. Mater. Sci.
  Lett.} \textbf{1986}, \emph{5}, 912--914\relax
\mciteBstWouldAddEndPuncttrue
\mciteSetBstMidEndSepPunct{\mcitedefaultmidpunct}
{\mcitedefaultendpunct}{\mcitedefaultseppunct}\relax
\EndOfBibitem
\bibitem[Welzmiller \latin{et~al.}(2014)Welzmiller, Hennersdorf, Fitch, and
  Oeckler]{Welzmiller2014}
Welzmiller,~S.; Hennersdorf,~F.; Fitch,~A.; Oeckler,~O. {Solid solution series
  between CdIn$_2$Te$_4$ and AgInTe$_2$ investigated by resonant X-ray
  scattering}. \emph{Z. Anorg. Allg. Chem.} \textbf{2014}, \emph{640},
  3135--3142\relax
\mciteBstWouldAddEndPuncttrue
\mciteSetBstMidEndSepPunct{\mcitedefaultmidpunct}
{\mcitedefaultendpunct}{\mcitedefaultseppunct}\relax
\EndOfBibitem
\bibitem[Welzmiller \latin{et~al.}(2015)Welzmiller, Hennersdorf, Schlegel,
  Fitch, Wagner, and Oeckler]{Welzmiller2015}
Welzmiller,~S.; Hennersdorf,~F.; Schlegel,~R.; Fitch,~A.; Wagner,~G.;
  Oeckler,~O. Silver Indium Telluride Semiconductors and Their Solid Solutions
  with Cadmium Indium Telluride: Structure and Physical Properties.
  \emph{Inorg. Chem.} \textbf{2015}, \emph{54}, 5745--5756, PMID:
  26023890\relax
\mciteBstWouldAddEndPuncttrue
\mciteSetBstMidEndSepPunct{\mcitedefaultmidpunct}
{\mcitedefaultendpunct}{\mcitedefaultseppunct}\relax
\EndOfBibitem
\bibitem[Range \latin{et~al.}(1969)Range, Engert, and Weiss]{Range1969a}
Range,~K.-J.; Engert,~G.; Weiss,~A. High pressure transformations of ternary
  chalcogenides with chalcopyrite structure — I. Indium-containing compounds.
  \emph{Solid State Commun.} \textbf{1969}, \emph{7}, 1749--1752\relax
\mciteBstWouldAddEndPuncttrue
\mciteSetBstMidEndSepPunct{\mcitedefaultmidpunct}
{\mcitedefaultendpunct}{\mcitedefaultseppunct}\relax
\EndOfBibitem
\bibitem[Range \latin{et~al.}(1969)Range, Engert, and Weiss]{Range1969b}
Range,~K.-J.; Engert,~G.; Weiss,~A. AgInTe$_2$-III — eine metastabile
  Modifikation des AgInTe$_2$? \emph{Z. Naturforsch. B} \textbf{1969},
  \emph{24}, 813--814\relax
\mciteBstWouldAddEndPuncttrue
\mciteSetBstMidEndSepPunct{\mcitedefaultmidpunct}
{\mcitedefaultendpunct}{\mcitedefaultseppunct}\relax
\EndOfBibitem
\bibitem[Bovornratanaraks \latin{et~al.}(2010)Bovornratanaraks, Kotmool,
  Yoodee, McMahon, and Ruffolo]{Bovornratanaraks2010}
Bovornratanaraks,~T.; Kotmool,~K.; Yoodee,~K.; McMahon,~M.~I.; Ruffolo,~D. High
  pressure structural studies of {AgInTe}$_2$. \emph{J. Phys.: Conf. Ser.}
  \textbf{2010}, \emph{215}, 012008\relax
\mciteBstWouldAddEndPuncttrue
\mciteSetBstMidEndSepPunct{\mcitedefaultmidpunct}
{\mcitedefaultendpunct}{\mcitedefaultseppunct}\relax
\EndOfBibitem
\bibitem[Kotmool \latin{et~al.}(2015)Kotmool, Bovornratanaraks, and
  Yoodee]{Kotmool2015}
Kotmool,~K.; Bovornratanaraks,~T.; Yoodee,~K. Ab initio study of structural
  phase transformations and band gap of chalcopyrite phase in AgInTe$_2$ under
  high pressure. \emph{Solid State Commun.} \textbf{2015}, \emph{220},
  25--30\relax
\mciteBstWouldAddEndPuncttrue
\mciteSetBstMidEndSepPunct{\mcitedefaultmidpunct}
{\mcitedefaultendpunct}{\mcitedefaultseppunct}\relax
\EndOfBibitem
\bibitem[Lütgemeier(1964)]{Lutgemeier1964}
Lütgemeier,~H. Die chemische Verschiebung der Kernresonanzlinien in
  A$^{\mathrm{III}}$B$^{\mathrm{V}}$-Verbindungen. \emph{Z. Naturforsch. A}
  \textbf{1964}, \emph{19}, 1297--1300\relax
\mciteBstWouldAddEndPuncttrue
\mciteSetBstMidEndSepPunct{\mcitedefaultmidpunct}
{\mcitedefaultendpunct}{\mcitedefaultseppunct}\relax
\EndOfBibitem
\bibitem[Becker and Schäfgen(1979)Becker, and Schäfgen]{Becker1979}
Becker,~K.; Schäfgen,~H. NMR chemical shifts in copper (I) chalcogen
  compounds. \emph{Solid State Commun.} \textbf{1979}, \emph{32},
  1107--1109\relax
\mciteBstWouldAddEndPuncttrue
\mciteSetBstMidEndSepPunct{\mcitedefaultmidpunct}
{\mcitedefaultendpunct}{\mcitedefaultseppunct}\relax
\EndOfBibitem
\bibitem[Vanderah and Nissan(1988)Vanderah, and Nissan]{Vanderah1988}
Vanderah,~T.; Nissan,~R. $^{31}$P MAS NMR of a {II–IV–V}$_2$
  chalcopyrite-type series. \emph{J. Phys. Chem. Solids} \textbf{1988},
  \emph{49}, 1335--1338\relax
\mciteBstWouldAddEndPuncttrue
\mciteSetBstMidEndSepPunct{\mcitedefaultmidpunct}
{\mcitedefaultendpunct}{\mcitedefaultseppunct}\relax
\EndOfBibitem
\bibitem[Shulman \latin{et~al.}(1958)Shulman, Wyluda, and
  Hrostowski]{Shulman1958}
Shulman,~R.~G.; Wyluda,~B.~J.; Hrostowski,~H.~J. {Nuclear Magnetic Resonance in
  Semiconductors. III. Exchange Broadening in GaAs and InAs}. \emph{Phys. Rev.}
  \textbf{1958}, \emph{109}, 808--809\relax
\mciteBstWouldAddEndPuncttrue
\mciteSetBstMidEndSepPunct{\mcitedefaultmidpunct}
{\mcitedefaultendpunct}{\mcitedefaultseppunct}\relax
\EndOfBibitem
\bibitem[Engelsberg and Norberg(1972)Engelsberg, and Norberg]{Engelsberg1972}
Engelsberg,~M.; Norberg,~R.~E. Nuclear Magnetic Resonance and Nuclear-Spin
  Dynamics in InP. \emph{Phys. Rev. B} \textbf{1972}, \emph{5},
  3395--3406\relax
\mciteBstWouldAddEndPuncttrue
\mciteSetBstMidEndSepPunct{\mcitedefaultmidpunct}
{\mcitedefaultendpunct}{\mcitedefaultseppunct}\relax
\EndOfBibitem
\bibitem[Han \latin{et~al.}(1988)Han, Timken, and Oldfield]{Han1988}
Han,~O.~H.; Timken,~H. K.~C.; Oldfield,~E. {Solid‐state
  ‘‘magic‐angle’’ sample‐spinning nuclear magnetic resonance
  spectroscopic study of group III–V (13–15) semiconductors}. \emph{J.
  Chem. Phys.} \textbf{1988}, \emph{89}, 6046--6052\relax
\mciteBstWouldAddEndPuncttrue
\mciteSetBstMidEndSepPunct{\mcitedefaultmidpunct}
{\mcitedefaultendpunct}{\mcitedefaultseppunct}\relax
\EndOfBibitem
\bibitem[Tomaselli \latin{et~al.}(1998)Tomaselli, deGraw, Yarger, Augustine,
  and Pines]{Tomaselli1998}
Tomaselli,~M.; deGraw,~D.; Yarger,~J.~L.; Augustine,~M.~P.; Pines,~A. Scalar
  and anisotropic $J$ interactions in undoped InP: A triple-resonance NMR
  study. \emph{Phys. Rev. B} \textbf{1998}, \emph{58}, 8627--8633\relax
\mciteBstWouldAddEndPuncttrue
\mciteSetBstMidEndSepPunct{\mcitedefaultmidpunct}
{\mcitedefaultendpunct}{\mcitedefaultseppunct}\relax
\EndOfBibitem
\bibitem[Adolphi \latin{et~al.}(1992)Adolphi, Conradi, and Buhro]{Adolphi1992}
Adolphi,~N.~L.; Conradi,~M.~S.; Buhro,~W.~E. {The $^{31}$P NMR spectrum of
  InP}. \emph{J. Phys. Chem. Solids} \textbf{1992}, \emph{53}, 1073--1074\relax
\mciteBstWouldAddEndPuncttrue
\mciteSetBstMidEndSepPunct{\mcitedefaultmidpunct}
{\mcitedefaultendpunct}{\mcitedefaultseppunct}\relax
\EndOfBibitem
\bibitem[Iijima \latin{et~al.}(2004)Iijima, Hashi, Goto, Shimizu, and
  Ohki]{Iijima2004}
Iijima,~T.; Hashi,~K.; Goto,~A.; Shimizu,~T.; Ohki,~S. {Indirect nuclear
  spin–spin coupling in InP studied by CP/MAS NMR}. \emph{Physica B}
  \textbf{2004}, \emph{346-347}, 476--478, Proceedings of the 7th International
  Symposium on Research in High Magnetic Fields\relax
\mciteBstWouldAddEndPuncttrue
\mciteSetBstMidEndSepPunct{\mcitedefaultmidpunct}
{\mcitedefaultendpunct}{\mcitedefaultseppunct}\relax
\EndOfBibitem
\bibitem[Iijima \latin{et~al.}(2006)Iijima, Hashi, Goto, Shimizu, and
  Ohki]{Iijima2006}
Iijima,~T.; Hashi,~K.; Goto,~A.; Shimizu,~T.; Ohki,~S. {Anisotropic indirect
  nuclear spin–spin coupling in InP: $^{31}$P CP NMR study under slow MAS
  condition}. \emph{Chem. Phys. Lett.} \textbf{2006}, \emph{419}, 28--32\relax
\mciteBstWouldAddEndPuncttrue
\mciteSetBstMidEndSepPunct{\mcitedefaultmidpunct}
{\mcitedefaultendpunct}{\mcitedefaultseppunct}\relax
\EndOfBibitem
\bibitem[Haase \latin{et~al.}(2009)Haase, Goh, Meissner, Alireza, and
  Rybicki]{Haase2009}
Haase,~J.; Goh,~S.~K.; Meissner,~T.; Alireza,~P.~L.; Rybicki,~D. High
  sensitivity nuclear magnetic resonance probe for anvil cell pressure
  experiments. \emph{Rev. Sci. Instrum.} \textbf{2009}, \emph{80}, 073905\relax
\mciteBstWouldAddEndPuncttrue
\mciteSetBstMidEndSepPunct{\mcitedefaultmidpunct}
{\mcitedefaultendpunct}{\mcitedefaultseppunct}\relax
\EndOfBibitem
\bibitem[Meissner \latin{et~al.}(2013)Meissner, Goh, Haase, Richter, Koepernik,
  and Eschrig]{Meissner2013}
Meissner,~T.; Goh,~S.~K.; Haase,~J.; Richter,~M.; Koepernik,~K.; Eschrig,~H.
  Nuclear magnetic resonance at up to 10.1 {GPa} pressure detects an electronic
  topological transition in aluminum metal. \emph{J. Phys.: Condens. Matter}
  \textbf{2013}, \emph{26}, 015501\relax
\mciteBstWouldAddEndPuncttrue
\mciteSetBstMidEndSepPunct{\mcitedefaultmidpunct}
{\mcitedefaultendpunct}{\mcitedefaultseppunct}\relax
\EndOfBibitem
\bibitem[Kattinger \latin{et~al.}(2021)Kattinger, Guehne, Tsankov, Jurkutat,
  Erb, and Haase]{Kattinger2021}
Kattinger,~C.; Guehne,~R.; Tsankov,~S.; Jurkutat,~M.; Erb,~A.; Haase,~J.
  {Moissanite anvil cell single crystal NMR at pressures of up to 4.4 GPa}.
  \emph{Rev. Sci. Instrum.} \textbf{2021}, \emph{92}, 113901\relax
\mciteBstWouldAddEndPuncttrue
\mciteSetBstMidEndSepPunct{\mcitedefaultmidpunct}
{\mcitedefaultendpunct}{\mcitedefaultseppunct}\relax
\EndOfBibitem
\bibitem[Meier \latin{et~al.}(2015)Meier, Reichardt, and Haase]{Meier2015}
Meier,~T.; Reichardt,~S.; Haase,~J. High-sensitivity NMR beyond 200,000
  atmospheres of pressure. \emph{J. Magn. Reson.} \textbf{2015}, \emph{257},
  39--44\relax
\mciteBstWouldAddEndPuncttrue
\mciteSetBstMidEndSepPunct{\mcitedefaultmidpunct}
{\mcitedefaultendpunct}{\mcitedefaultseppunct}\relax
\EndOfBibitem
\bibitem[Harris \latin{et~al.}(2008)Harris, Becker, De~Menezes, Granger,
  Hoffman, and Zilm]{Harris2008}
Harris,~R.~K.; Becker,~E.~D.; De~Menezes,~S. M.~C.; Granger,~P.;
  Hoffman,~R.~E.; Zilm,~K.~W. {Further conventions for NMR shielding and
  chemical shifts (IUPAC recommendations 2008)}. \emph{Mag. Res. Chem.}
  \textbf{2008}, \emph{46}, 582--598\relax
\mciteBstWouldAddEndPuncttrue
\mciteSetBstMidEndSepPunct{\mcitedefaultmidpunct}
{\mcitedefaultendpunct}{\mcitedefaultseppunct}\relax
\EndOfBibitem
\bibitem[Meissner \latin{et~al.}(2010)Meissner, Goh, Haase, Meier, Rybicki, and
  Alireza]{Meissner2010}
Meissner,~T.; Goh,~S.~K.; Haase,~J.; Meier,~B.; Rybicki,~D.; Alireza,~P. New
  Approach to High-Pressure Nuclear Magnetic Resonance with Anvil Cells.
  \emph{J. Low Temp. Phys.} \textbf{2010}, \emph{159}, 284\relax
\mciteBstWouldAddEndPuncttrue
\mciteSetBstMidEndSepPunct{\mcitedefaultmidpunct}
{\mcitedefaultendpunct}{\mcitedefaultseppunct}\relax
\EndOfBibitem
\bibitem[Forman \latin{et~al.}(1972)Forman, Piermarini, Barnett, and
  Block]{Forman1972}
Forman,~R.~A.; Piermarini,~G.~J.; Barnett,~J.~D.; Block,~S. {Pressure
  measurement made by the utilization of ruby sharp-line luminescence}.
  \emph{Science} \textbf{1972}, \emph{176}, 284--285\relax
\mciteBstWouldAddEndPuncttrue
\mciteSetBstMidEndSepPunct{\mcitedefaultmidpunct}
{\mcitedefaultendpunct}{\mcitedefaultseppunct}\relax
\EndOfBibitem
\bibitem[Ruiz-Morales \latin{et~al.}(1997)Ruiz-Morales, Schreckenbach, and
  Ziegler]{Morales1997}
Ruiz-Morales,~Y.; Schreckenbach,~G.; Ziegler,~T. Calculation of $^{125}$Te
  Chemical Shifts Using Gauge-Including Atomic Orbitals and Density Functional
  Theory. \emph{J. Phys. Chem. A} \textbf{1997}, \emph{101}, 4121--4127\relax
\mciteBstWouldAddEndPuncttrue
\mciteSetBstMidEndSepPunct{\mcitedefaultmidpunct}
{\mcitedefaultendpunct}{\mcitedefaultseppunct}\relax
\EndOfBibitem
\bibitem[sol()]{solid}
For quadrupole echoes applied on macroscopic samples and coils, we used a
  static probe with a low-Q RF circuit so that we could use short excitation
  pulses of 1$\mu$s. While this ensured small spectral distortions (as we
  verified by using large frequency offsets) we could not avoid the formation
  of significant Solomon echoes from multiple quantum transitions (powder
  sample). The latter interfere with recording of single quantum transition
  quadrupole echoes and lead to severe spectral distortions, unless we use long
  echo delays. Unfortunately, due to the spin echo decay most of the signal has
  decayed at sufficiently long delays so that the thus obtained spectra are
  compromised, as well. Nevertheless, within error, we find a similar
  featureless line shape as for the frequency stepped spin echoes, while we
  also ascertained that the small difference is not due to a bandwidth
  limitation.\relax
\mciteBstWouldAddEndPunctfalse
\mciteSetBstMidEndSepPunct{\mcitedefaultmidpunct}
{}{\mcitedefaultseppunct}\relax
\EndOfBibitem
\bibitem[Vleck(1948)]{vVleck1948}
Vleck,~J.~V. The Dipolar Broadening of Magnetic Resonance Lines in Crystals.
  \emph{Phys. Rev.} \textbf{1948}, 116–1183\relax
\mciteBstWouldAddEndPuncttrue
\mciteSetBstMidEndSepPunct{\mcitedefaultmidpunct}
{\mcitedefaultendpunct}{\mcitedefaultseppunct}\relax
\EndOfBibitem
\bibitem[Aikebaier \latin{et~al.}(2012)Aikebaier, Kurosaki, Sugahara, Ohishi,
  Muta, and Yamanaka]{Aikebaier2012}
Aikebaier,~Y.; Kurosaki,~K.; Sugahara,~T.; Ohishi,~Y.; Muta,~H.; Yamanaka,~S.
  High-temperature thermoelectric properties of non-stoichiometric
  Ag$_{1-x}$InTe$_2$ with chalcopyrite structure. \emph{Mater. Sci. Eng., B}
  \textbf{2012}, \emph{122}, 999--1002\relax
\mciteBstWouldAddEndPuncttrue
\mciteSetBstMidEndSepPunct{\mcitedefaultmidpunct}
{\mcitedefaultendpunct}{\mcitedefaultseppunct}\relax
\EndOfBibitem
\bibitem[Van~Kranendonk(1954)]{Kranendonk1954}
Van~Kranendonk,~J. Theory of quadrupolar nuclear spin-lattice relaxation.
  \emph{Physica} \textbf{1954}, \emph{20}, 781 -- 800\relax
\mciteBstWouldAddEndPuncttrue
\mciteSetBstMidEndSepPunct{\mcitedefaultmidpunct}
{\mcitedefaultendpunct}{\mcitedefaultseppunct}\relax
\EndOfBibitem
\bibitem[Van~Kranendonk and Walker(1967)Van~Kranendonk, and
  Walker]{Kranendonk1967}
Van~Kranendonk,~J.; Walker,~M. {Theory of quadrupolar nuclear spin-lattice
  relaxation due to anharmonic Raman phonon processes}. \emph{Phys. Rev. Lett.}
  \textbf{1967}, \emph{18}, 701--703\relax
\mciteBstWouldAddEndPuncttrue
\mciteSetBstMidEndSepPunct{\mcitedefaultmidpunct}
{\mcitedefaultendpunct}{\mcitedefaultseppunct}\relax
\EndOfBibitem
\bibitem[Haase and Oldfield(1993)Haase, and Oldfield]{Haase1993b}
Haase,~J.; Oldfield,~E. {Spin-Echo Behavior of Nonintegral-Spin Quadrupolar
  Nuclei in Inorganic Solids}. \emph{J. Magn. Reson. Ser. A} \textbf{1993},
  \emph{101}, 30--40\relax
\mciteBstWouldAddEndPuncttrue
\mciteSetBstMidEndSepPunct{\mcitedefaultmidpunct}
{\mcitedefaultendpunct}{\mcitedefaultseppunct}\relax
\EndOfBibitem
\end{mcitethebibliography}

\end{document}